
%
%
\hyphenation{Pij-pers}
\hyphenation{mode-set}
%
\newcount\eqnumber
\eqnumber=1
\def\neqn{{\rm(\the\eqnumber)}\global\advance\eqnumber by 1}
\def\refeq#1){\advance\eqnumber by -#1 {\rm(\the\eqnumber)} \advance
\eqnumber by #1}
\def\eqnam#1#2{\immediate\write1{\xdef\ #2{(\the\eqnumber}}\xdef#1{(\the\eqnumber}}
\newcount\fignumber
\fignumber=1
\def\nfig{\global\advance\fignumber by 1}
\def\refig#1{\advance\fignumber by -#1 \the\fignumber \advance\fignumber by #1}
\def\fignam#1#2{\immediate\write1{\xdef\ #2{\the\fignumber}}\xdef#1{\the\fignumber}}
\def\note #1]{{\bf #1]}}
\def\draft{\headline{\bf File: \jobname\hfill DRAFT\hfill\today}}
\def\ref{\par\noindent
	\hangindent=0.7 truecm
	\hangafter=1}
%
%

\draft

\MAINTITLE{Research Note : Solar rotation inversions and the relationship
between a-coefficients and mode splittings }


\AUTHOR={ F.P. Pijpers }

\OFFPRINTS{F.P. Pijpers}

\INSTITUTE{
Theoretical Astrophysics Center, Institute for Physics and Astronomy, 
Aarhus University, Ny Munkegade, 8000~Aarhus~C, Denmark }

\DATE{Received ; accepted}
\ABSTRACT{ From the observing campaigns of a number of helioseismic 
telescope networks such as the Global Oscillation Network Group (GONG) and 
also from the Solar Heliospheric Observatory satellite (SoHO), 
helioseismologists now have data on in excess of 20000 oscillation 
frequencies $\nu$ of the modes known collectively as the 5 minute oscillations.
Depending on the order and degree of the modes these data are available
either as individual frequencies of modes with known radial order $n$,
spherical harmonic degree $l$, and order $m$, or as so-called `a-coefficients'
which are obtained by fitting polynomials to ridges of power in the
$(l,\nu )$ diagram. Although there is a one-to-one relationship between 
these two types of data, the conversion between them is not straightforward
which complicates comparison of helioseismic inference between the two types
of data. Their relationship with each other and with the solar rotation rate
is clarified in this paper.
}


\THESAURUS{09(06.18.2 - 06.09.1 - 03.13.1)}

\maketitle

\titlea{Introduction}

For observations that are well resolved in space and in time the 
oscillations of the Sun can be projected uniquely onto its pulsation 
eigenmodes, which are products of functions of radius and
of spherical harmonic functions. Each mode, and therefore each measured
oscillation frequency, is uniquely identified by three numbers~: the radial
order $n$, and the degree $l$ and the azimuthal order $m$ of the spherical 
harmonic. For a non-rotating star the frequency is degenerate with respect 
to $m$. Rotation removes this degeneracy and the frequencies are split 
multiplets. Individual mode splittings $D_{nlm}$ ($m>0$) are related to 
the mode frequencies by~:
$$
D_{nlm} = {\nu_{nlm} - \nu_{nl\,-m}\over 2 m}
\eqno\neqn
$$
and to the solar rotation rate $\Omega(r, \theta )$ by~:
\eqnam\Drotrel{Drotrel}
$$
2\pi D_{nlm}\ =\ \int\limits_0^1 \int\limits_{-1}^1\, {\rm d} r\, 
{\rm d}\cos \theta\, K_{nlm}(r,\theta) \Omega(r,\theta) 
\eqno\neqn
$$
where $r$ is the fractional radius and $\theta$ the colatitude. The 
$K_{nlm}$ are the mode kernels for rotation given by~:
\eqnam{\Separable}{Separable}
$$
K_{nlm} (r, \theta)\ \equiv\ F_1^{nl}(r) G_1^{lm}(\theta)\;+\;
F_2^{nl}(r) G_2^{lm}(\theta)
\eqno\neqn
$$
where
\eqnam{\Fodef}{Fodef}
$$
F^{nl}_1\ \equiv\ \rho(r) r^2 \left[ \xi_{nl}^2(r) - 2L^{-1} \xi_{nl}(r)
\eta_{nl} (r) + \eta_{nl}^2(r) \right] / I_{nl}
\eqno\neqn
$$
and
\eqnam{\Ftdef}{Ftdef}
$$
F^{nl}_2\ \equiv\ \rho(r) r^2 \left[ L^{-2} \eta_{nl}^2(r) \right]
/ I_{nl}
\eqno\neqn
$$
Here $\rho$ is the density as a function of radius, $L^2 \equiv l(l+1)$ 
and $\xi_{nl}$, $\eta_{nl}$ are the radial and horizontal components of 
the displacement eigenfunction in the non-rotating star, and
$$
I_{nl}\ =\ \int\limits_0^1 {\rm d} r\, \rho r^2 (\xi_{nl}^2 +
\eta_{nl}^2 )
\eqno\neqn
$$
Further (using $u\ \equiv\ \cos\theta $)~:
\eqnam{\Godef}{Godef}
$$
G^{lm}_1\ \equiv\ { (l - \vert m \vert)! \over (l + \vert m \vert)!} (l+1/2)
\left[ P_l^m(u) \right]^2
\eqno\neqn
$$
\eqnam{\Gtdef}{Gtdef}
$$
\eqalign{
G^{lm}_2\ \equiv\ { (l - \vert m \vert)! \over (l + \vert m \vert)!} (l+1/2)
&\left[ (1-u^2)\left({{\rm d} P_l^m(u)\over{\rm d} u}\right)^2 + \right. \cr
2u P_l^m(u) {{\rm d} P_l^m(u)\over {\rm d} u} + 
&\left.\left({m^2\over 1-u^2}-L^2 \right) \left(P_l^m(u) 
\right)^2\right] \cr}
\eqno\neqn
$$
Here the $P_l^m(u)$ are associated Legendre polynomials. 
This separation of $K_{nlm}$ into $F^{nl}_1, F^{nl}_2$ and $G^{lm}_1, G^{lm}_2$ 
is not unique. Many other choices are possible but this choice is particularly
useful for the purposes of this paper. The same choice is made in 
Pijpers \& Thompson (1996). One advantage of this choice is that~:
\eqnam{\Gfunrel}{Gfunrel}
$$
G^{lm}_2\ =\ {1\over 2} (1-u^2) {{\rm d}^2 G^{lm}_1\over{\rm d} u^2}
\eqno\neqn
$$
Note that with the definitions \Godef), \Gtdef) of $G_{1,2}^{lm}$ the 
normalization is such that~:
\eqnam\GLMnor{GLMnor}
$$
\int\limits_{-1}^{1} {\rm d} u\, G_1^{lm}(u) = 1 = - \int\limits_{-1}^{1} 
{\rm d} u\, G_2^{lm}(u)
\eqno\neqn
$$

\titlea{The a-coefficient fitting}

The a-coefficients are constructed by fitting polynomials in $m$ to 
ridges of power in the $(l, \nu)$ diagram. Details of their use can be 
found in eg. Schou et al. (1994, hereafter SCDT). The relationship 
between a-coefficients and mode frequencies is~:
$$
\nu_{nlm} = \sum_{s=0}^{s_{\rm max}} a_{nl\, s} {\cal P}_s^{(l)} (m)
\eqno\neqn
$$
Here the polynomials ${\cal P}_s^{(l)}$ have degree $s$ and are often 
normalized so that ${\cal P}_s^{(l)} (l) = l$. As noted in SCDT, who follow 
the projection proposed by Ritzwoller \& Lavely (1991), these properties 
together with the orthogonality condition~:
$$
\sum_{m=-l}^l {\cal P}_s^{(l)} (m) {\cal P}_{s'}^{(l)} (m) =0 \quad 
{\rm for}\ s\neq s'
\eqno\neqn
$$
specify the polynomials ${\cal P}_s^{(l)}$ completely. As noted in
Appendix A of SCDT the polynomials are usually constructed in an iterative 
process.

In practice it can be impossible, eg. for high values of $l$, to identify 
the individual peaks of power of every mode, in which case the fitting 
must proceed using a slightly different scheme than that described below.
However at least formally the fitting always can be described as a least 
squares fit of polynomials to individual frequencies.
In such a least-squares fit for the a-coefficients the $\chi^2$-measure to
be minimized is~:
\eqnam\LSmipo{LSmipo}
$$
\chi^2 = \sum_{m=-l}^l \left ( \nu_{nlm} - \sum_{s=0}^{s_{\rm max}} a_{nl\, s}
{\cal P}_s^{(l)} (m) \right )^2
\eqno\neqn
$$
Differentiation of Eq. \LSmipo) with respect to $a_{nl\, s}$ yields~:
$$
0=\sum_{m=-l}^l \left ( \nu_{nlm} - \sum_{s=0}^{s_{\rm max}} a_{nl\, s} 
{\cal P}_s^{(l)} (m) \right ) {\cal P}_{s'}^{(l)} (m)
\eqno\neqn
$$
or~:
$$
\eqalign{
\sum_{m=-l}^l \nu_{nlm} {\cal P}_{s'}^{(l)} (m) &=
\sum_{m=-l}^l \sum_{s=0}^{s_{\rm max}} a_{nl\, s} {\cal P}_s^{(l)} (m) 
{\cal P}_{s'}^{(l)} (m) \cr 
&= \sum_{s=0}^{s_{\rm max}} a_{nl\, s} \sum_{m=-l}^l {\cal P}_s^{(l)} (m) 
{\cal P}_{s'}^{(l)} (m)
\cr}
\eqno\neqn
$$
and using the orthogonality of the ${\cal P}$'s then leads to~:
$$
\sum_{m=-l}^l \nu_{nlm} {\cal P}_s^{(l)} (m) =
a_{nl\, s} \sum_{m=-l}^l \left ({\cal P}_s^{(l)} (m) \right )^2
\eqno\neqn
$$
from which is obtained~:
$$
a_{nl\, s} = {{\sum_{m=-l}^l \nu_{nlm} {\cal P}_s^{(l)} (m)} \over
{\sum_{m^\prime=-l}^l \left ({\cal P}_s^{(l)} (m^\prime) \right )^2}} \equiv 
\sum_{m=-l}^l c^{(s)}_{lm}\nu_{nlm}
\eqno\neqn
$$
where the $c^{(s)}_{lm}$ are defined as~:
$$
c^{(s)}_{lm} \equiv {{{\cal P}_s^{(l)} (m)} \over {\sum_{m^\prime=-l}^l \left
({\cal P}_s^{(l)} (m^\prime) \right )^2}}
\eqno\neqn
$$
In the case of inversion for the determination of the solar rotation rate 
only the a-coefficients for odd $s$ are of interest since these are a 
measure of the flow that is symmetric between northern and southern 
hemispheres. Since the coefficients satisfy 
$c^{(2s+1)}_{l\, -m} = -c^{(2s+1)}_{lm}$ (i.e. ${\cal P}_{2s+1}^{(l)} 
(-m) = -{\cal P}_{2s+1}^{(l)} (m)$), it is clear that linear coefficients 
$\gamma_{2s+1}^m$ exist such that~:
\eqnam\asrel{asrel}
$$
a_{nl\, 2s+1}\ =\ \sum\limits_{m=1}^{l} \gamma^{m}_{2s+1} D_{nlm}
\eqno\neqn
$$
and they can be obtained from the polynomials ${\cal P}_s^{(l)}$ by using~:
\eqnam\gamPrel{gamPrel}
$$
\gamma_{2s+1}^m =  {m {{\cal P}_{2s+1}^{(l)} (m)} \over 
{\sum_{m'=1}^l \left ({\cal P}_{2s+1}^{(l)} (m^\prime) \right )^2}}
\eqno\neqn
$$

\titlea{The projection of $\Omega$}

As was demonstrated by Ritzwoller \& Lavely (1991) and SCDT, this 
choice of orthogonal fitting polynomials ${\cal P}_s$ leads to $a_s$
such that they correspond one-to-one to a projection of $\Omega$ onto
polynomials $W_s$ (i.e. Eq. (3.21) of SCDT)~:
$$
2\pi a_{nl\, 2s+1}\ =\ \int\limits_{0}^{1} {\rm d} r\, K^{(a)}_{nl\, s}(r) 
\Omega_s(r) 
\eqno\neqn
$$
so that, if the $\Omega_s (r)$ are reconstructed from a sequence of
1-dimensional inversions using the a-coefficients, $\Omega (r,\theta)$ 
can be reconstructed from~:
\eqnam\OmegOneTwo{OmegOneTwo}
$$
\Omega (r,\theta) = \sum\limits_{s=o}^{s_{\rm max}} \Omega_s(r) W_s (\theta)
\eqno\neqn
$$
Since the a-coefficients are independent the associated projection functions
$W_s$ must be orthogonal to the latitudinal response functions associated
with the a-coeffi\-cients. Combination of Eqs. \Drotrel), \asrel) and 
\OmegOneTwo) yields the following orthogonality conditions~:
\eqnam\WGorth{WGorth}
$$
\eqalign{
\int\limits_{-1}^{1} {\rm d}\cos\theta\; \sum\limits_{m=1}^{l}
\gamma^{m}_{2s+1} G^{lm}_1 (\theta) W_{s'} (\theta) &= g_{1\, s} \delta_{s\, s'} 
\cr
\int\limits_{-1}^{1} {\rm d}\cos\theta\; \sum\limits_{m=1}^{l}
\gamma^{m}_{2s+1} G^{lm}_2 (\theta) W_{s'} (\theta) &= g_{2\, s} \delta_{s\, s'} 
\cr}
\eqno\neqn
$$
so that the kernels $K^{(a)}_{nl\, s}(r)$ are given by~:
$$
K^{(a)}_{nl\, s} (r)\ \equiv\ F_1^{nl}(r) g_{1\, s}\; +\;
F_2^{nl}(r) g_{2\, s}
\eqno\neqn
$$
If conditions \WGorth) are not satisfied there is `cross-talk' between 
a-coefficients for different $s$. Since the set $W_s$ is itself 
complete it follows from Eq. \WGorth) that the linear sums of the $G_1$ 
and of the $G_2$ must be identical apart from a constant factor~:
\eqnam\Gadef{Gadef}
$$
\eqalign{
\sum\limits_{m=1}^{l} \gamma^{m}_{2s+1} G^{lm}_2 \equiv &G^{(l)\, 2s+1}_2 
= \cr &\lambda G^{(l)\, 2s+1}_1 \equiv \lambda \sum\limits_{m=1}^{l}
\gamma^{m}_{2s+1} G^{lm}_1 \cr}
\eqno\neqn
$$
Combination of Eqs. \Gfunrel) and \Gadef) yields the equation~:
\eqnam\GoneDE{GoneDE}
$$
{1\over 2} (1-u^2) {{\rm d}^2 G^{(l)}_1\over{\rm d} u^2} = \lambda G^{(l)}_1
\eqno\neqn
$$
Integrating the terms in Eq. \Gadef) over $u$ and using Eq. \GLMnor) 
shows that~:
\eqnam\Wcondit{Wcondit}
$$
\lambda = -1  \quad {\rm or}\quad \sum\limits_{m=1}^{l} \gamma_{2s+1}^m = 0
\eqno\neqn
$$
The regular solution of Eq. \GoneDE) is~:
$$
G^{(l)\, t}_1 = A^{(l)}_t (1-u^2) {{\rm d} P_t^0 (u) \over{\rm d} u}
\eqno\neqn
$$
for $\lambda = -{1\over 2} t (t+1) $.
It is convenient to choose $A^{(l)}_t$ such that~:
\eqnam\Acst{Acst}
$$
A^{(l)}_t = {(2t+1) v^{(l)}_{t}\over 2 t (t+1)} 
\eqno\neqn
$$
so that the latitudinal functions ($t=2s+1$) are~:
\eqnam\NLGa{NLGa}
$$
\eqalign{
G_1^{(l)\, 2s+1}(u) &= {1\over 2}{(4s+3) v^{(l)}_{2s+1}\over (2s+2)(2s+1)}
(1-u^2) {{\rm d}\over{\rm d} u} P_{2s+1}^0 (u) \cr
G_2^{(l)\, 2s+1}(u) &= -{1\over 2} (2s+1) (2s+2) G_1^{(l)\, 2s+1} \cr
&= -{1\over 4} (4s+3) v^{(l)}_{2s+1}(1-u^2) {{\rm d}\over{\rm d} u} 
P_{2s+1}^0 (u)
\cr}
\eqno\neqn
$$
With the definition of $G_1$ and $G_2$ through expressions \NLGa) 
it is simple to show that~:
\eqnam\GSnor{GSnor}
$$
\int\limits_{-1}^{1}{\rm d} u\, G_1^{(l)\, 2s+1}(u) = 
- \int\limits_{-1}^{1}{\rm d} u\, G_2^{(l)\, 2s+1}(u) =v^{(l)}_{2s+1} 
\delta_{s0}
\eqno\neqn
$$
By integrating all terms over $u$ in Eq. \Gadef), and by using Eqs. 
\GLMnor), \Gadef) and \GSnor), it follows that for all $s\neq 0$ the 
coefficients $\gamma$ satisfy $\sum_{m=1}^{l} \gamma_{2s+1}^m = 0$ as 
demanded by \Wcondit). The $v^{(l)}_{2s+1}$ are determined by the 
normalization of the fitting functions ${\cal P}$ which is demonstrated 
in Sect. 4. Equations \NLGa) can also be written as~:
\eqnam\NLGat{NLGat}
$$
\eqalign{
G_1^{(l)\, 2s+1}(u) &= -{1\over 2}{(4s+3) v^{(l)}_{2s+1}\over (2s+2)(2s+1)} 
(1-u^2)^{1\over 2} P_{2s+1}^1 (u) \cr
G_2^{(l)\, 2s+1}(u) &= {1\over 4}v^{(l)}_{2s+1} (4s+3) (1-u^2)^{1\over 2}
P_{2s+1}^1 (u) \cr}
\eqno\neqn$$
From the orthogonality properties of associated Legendre polynomials 
(cf. Gradshteyn \& Ryzhik, 1994) it follows that a natural choice for 
the projection polynomials $W_s$, independent of $l$, for $\Omega$ is~:
\eqnam\Wfifunc{Wfifunc}
$$
W_s (u) = -(1-u^2)^{-1/2} P_{2s+1}^1 (u)
\eqno\neqn
$$
From Eq. \Wfifunc) it follows that then conditions \WGorth) are 
satisfied with~: 
$$
\eqalign{
g_{1\, s} &= v^{(l)}_{2s+1} \cr 
g_{2\, s} &= -{1\over 2} v^{(l)}_{2s+1} (2s+1)(2s+2)\cr}
\eqno\neqn$$

\titlea{Normalization of the kernels}

The freedom of choosing the factors $v^{(l)}_{2s+1}$ introduced in 
Eq. \Acst) is directly related to the freedom of choosing the 
normalization of the fitting functions ${\cal P}_s$. Combination of
expressions \Gadef) and \NLGat) yields~:
\eqnam\SumgamG{SumgamG}
$$
\sum\limits_{m=1}^{l} \gamma^{m}_{2s+1} G^{lm}_1 = -{1\over 2}{(4s+3) 
v^{(l)}_{2s+1}\over (2s+2)(2s+1)} (1-u^2)^{1\over 2} P_{2s+1}^1 
\eqno\neqn$$
Multiplying Eq. \SumgamG) by $W_{s'}$ and integrating over $u$ on both 
sides yields~:
$$
\eqalign{
\sum\limits_{m=1}^{l} \gamma^{m}_{2s+1} \beta_{2s'+1}^{lm} \equiv 
\sum\limits_{m=1}^{l} \gamma^{m}_{2s+1} \int\limits_{-1}^1 {\rm d}u\, 
G^{lm}_1 &(u) W_{s'} (u) \cr &= \delta_{ss'}v_{2s+1}^{(l)} \cr}
\eqno\neqn$$
where the factors $\beta_{2s+1}^{lm}$ are introduced as shorthand 
notation for the integral. Using Eq. \gamPrel) it can be seen that~:
$$
\sum\limits_{m=1}^{l} m {\cal P}^{(l)}_{2s+1} (m) \beta_{2s+1}^{lm} =
\sum\limits_{m=1}^{l} \left( {\cal P}^{(l)}_{2s+1} 
(m) \right)^2
\eqno\neqn$$
From which it follows, using the orthogonality of the ${\cal P}$ polynomials,
that~:
$$
\beta_{2s+1}^{lm} = {1\over m}{\cal P}^{(l)}_{2s+1} v_{2s+1}^{(l)}
\eqno\neqn$$
and therefore~:
\eqnam\ClFoPs{ClFoPs}
$$
\eqalign{
{\cal P}^{(l)}_{2s+1} (m) &=  m \int\limits_{-1}^1 {\rm d}u\, 
G^{lm}_1 (u) W_s (u) \cr
&= {- m\over v^{(l)}_{2s+1}}{ (l - \vert m \vert)! \over (l + \vert m \vert)!} 
(l+1/2)\times \cr
&\hskip 1truecm\int\limits_{-1}^1 {\rm d}u\, (1-u^2)^{-1/2} 
P_{2s+1}^1 (u) \left[ P_l^m (u) \right]^2 \cr}
\eqno\neqn$$
Combination of the normalization condition ${\cal P}^{(l)}_{2s+1} (l) = l$ with 
Eq. \ClFoPs) yields~:
\eqnam\DiffInt{DiffInt}
$$
l = -{ l(2l+1)\over 2 v^{(l)}_{2s+1}}{\left[(2l-1)!!\right]^2\over 2l!} 
\int\limits_{-1}^1 {\rm d}u\, (1-u^2)^{l-1/2} P_{2s+1}^1 (u) 
\eqno\neqn$$
Equation \DiffInt) is derived by using the explicit expression for $P_l^l(u)$ 
(cf. Gradshteyn \& Ryzhik, 1994)~:
$$
P_l^l(u) = (-1)^l (2l-1)!! (1-u^2)^{l/2}
\eqno\neqn$$
The solution of the integral in Eq. \DiffInt) is (cf. Gradshteyn \& 
Ryzhik, 1994)~:
$$
\eqalign{
\int\limits_{-1}^1 {\rm d}u\, (1-u^2&)^{l-1/2} P_{2s+1}^1 (u) = \cr
&{2\pi\Gamma(l+1)\Gamma(l)\over\Gamma(l+s+{3\over 2})\Gamma(l-s)
\Gamma(s+1)\Gamma(-{1\over 2} -s ) } \cr}
\eqno\neqn$$
which means that the coefficients $v^{(l)}_{2s+1}$ must satisfy the 
following expression~:
\eqnam\vnorfac{vnorfac}
$$
\eqalign{
v^{(l)}_{2s+1} = (-1)^{s} {1\over l}&{(2l+1)! (2s+2)! (l+s+1)! \over s! 
(s+1)! (l-s-1)! (2l+2s+2)!} \cr
&\qquad\qquad\qquad {\rm for}\ s \leq l-1 \cr}
\eqno\neqn$$
For $s=0$ Eq. \vnorfac) reduces to $v^{(l)}_1 = 1$.
By using that $P_1^1(u) = -(1-u^2)^{1/2}$ it is easy to show that 
${\cal P}^{(l)}_1 (m) = m$ which is identical to the result obtained by 
SCDT in their Appendix A.

\titlea{Constructing the fitting functions}

In Appendix A of SCDT a relatively straightforward technique is
given for constructing the orthogonal fitting polynomials ${\cal P}.$
It is also possible to obtain the fitting functions by a different 
iteration scheme, taking Eq. \ClFoPs) as the starting point and making use 
of one of the functional relationships between Legendre polynomials of 
varying order and degree (cf. Gradshteyn \& Ryzhik, 1994)~:
\eqnam\PLfunrel{PLfunrel}
$$
P_{l-1}^{m+1} - P_{l+1}^{m+1} = (2l+1) \sqrt{1-u^2} P_l^m
\eqno\neqn$$
When this relationship is applied to $P_{2s+1}^{1}$ in Eq. \ClFoPs),
this yields~:
\eqnam\Piter{Piter}
$$
\eqalign{
{\cal P}^{(l)}_{2s+1} (m) &= {v^{(l)}_{2s-1}\over v^{(l)}_{2s+1}}
{\cal P}^{(l)}_{2s-1} (m) + { m(4s+1)\over v^{(l)}_{2s+1}}{(l-\vert m\vert)! 
\over (l + \vert m \vert)!} \times \cr
&(l+1/2) \int\limits_{-1}^1 {\rm d}u\, P_{2s}^0 (u) \left[ P_l^m (u) \right]^2 \cr}
\eqno\neqn$$
The integral can be evaluated by expanding the $P_{2s}$ in terms of 
power series of $(1-u^2)$ and making repeated use of Eq. \PLfunrel), 
now applied to the $P_l^m$. This leads to a weighted sum of a set of 
integrals of $[P_{l+k}^{m+\vert k\vert}]^2$ where k runs from $-s$ to 
$+s$. The integrals are easily evaluated from the normalization of 
associated Legendre polynomials. The evaluation of the weighting factors 
in terms of $k, l, m,$ and $s$ is rather cumbersome however, and there 
is a more convenient route to the same result. Integrals of products of 
three associated Legendre polynomials also occur regularly in quantum 
mechanics when adding angular momenta (cf. Merzbacher, 1970). It is 
straightforward to demonstrate that the integral in Eq. \Piter) can be 
written in terms of Clebsch-Gordan or Wigner coefficients, the construction 
schemes of which can be found in the literature (cf. Abramowitz \& Stegun). 
Equation \Piter) then becomes~:
$$
\eqalign{
{\cal P}^{(l)}_{2s+1} (m) = -&{s(2l+2s+1)\over (2s+1)(l-s)}
{\cal P}^{(l)}_{2s-1} (m) + {m(2l+1)\over v^{(l)}_{2s+1}}\times\cr
&\hskip 1.5cm\langle ll\, 00 \vert ll\, 2s\, 0\rangle\langle ll\, m\, -\kern-.3em m 
\vert ll \, 2s\, 0\rangle\cr}
\eqno\neqn$$
where the angular brackets are the usual notation for the Clebsch-Gordan 
coefficients.

\titlea{Implications for SOLA inversion methods}

All linear methods to perform inversions for rotation as a function of 
both radius and latitude reduce in algorithmic form to solving 
a large set of linear equations, ie. inverting a large matrix. The method 
of Subtractive Optimally Localized Averages (SOLA) (Pijpers \& Thompson, 
1994) is a possible inversion method which has proven to produce 
reliable results in other helioseismic problems.
In the SOLA method the (symmetric) matrix to be inverted has as its elements 
integrals over $r$ and $u$ of all the cross products of the individual 
kernels. For the 2-D inversion for solar rotation the number of elements 
in this matrix is of the order of $20000 \times 20000$ or more, which 
is prohibitive for the direct use of SOLA on this problem.
However, a direct consequence of obtaining the explicit expressions 
\NLGat) for the latitudinal kernels is that it can be demonstrated that 
this matrix has a sparse character. This can be used to make the direct 
use of SOLA quite feasible.

The matrix elements for the SOLA method are~:
\eqnam\Matels{Matels}
$$
{\cal C}^{js\, j's'} \equiv \int\limits_0^1 {\rm d}r \int\limits_{-1}^{1} 
{\rm d}u\ K_{nl\, s}(r,u) K_{n' l' \, s'} (r,u)
\eqno\neqn
$$
For convenience of notation the single multiplet index $j$ is introduced,
which runs over all combinations of the indices $n,l$.
The $K_{nl\, s}$ can be written as~:
\eqnam\KNLSdef{KNLSdef}
$$
K_{nl\, s}(r,u) = F_1^{nl} (r) G_1^{(l)\, 2s+1}(u) + F_2^{nl}(r) 
G_2^{(l)\, 2s+1} (u)
\eqno\neqn$$
Substituting Eq. \KNLSdef) into Eq. \Matels) for $s$ and for $s'$
leads to a sum of four integrals, and in each of the terms the 
integration over $u$ is the following~:
\eqnam\uintI{uintI}
$$
\int\limits_{-1}^1 {\rm d} u (4s+3)(1-u^2)^{1/2} P_{2s+1}^1 
(4s'+3)(1-u^2)^{1/2} P_{2s'+1}^1 
\eqno\neqn$$
Equation \uintI) is obtained by making use of the expressions \NLGat) for
the $G$ functions. Making use of Eq. \PLfunrel) this integral can be
rewritten as~:
\eqnam\Gacpi{Gacpi}
$$
\eqalign{
\int\limits_{-1}^1 {\rm d} u &\left[P_{2s}^2 -P_{2s+2}^2 \right]
\left[P_{2s'}^2 -P_{2s'+2}^2 \right]\ =\cr 
\delta_{s-1\, s'}&\left[{-2\over 4s+1 }{(2s+2)!\over(2s-2)!}\right] + \cr
\delta_{ss'}&\left[{2\over 4s+1 }{(2s+2)!\over(2s-2)!} + {2\over 4s+5 }
{(2s+4)!\over(2s)!}\right] + \cr
\delta_{s+1\, s'}&\left[{-2\over 4s+5 } {(2s+4)!\over(2s)!}\right]
\cr}
\eqno\neqn$$
The right-hand side of Eq. \Gacpi) is derived by making use of the 
orthogonality properties and normalization of associated Legendre 
polynomials (cf. Abramowitz \& Stegun, 1972), where the $\delta$ is the 
Kronecker symbol. From expression \Gacpi)
it follows that the integral of the cross-product of two mode kernels
is not equal to $0$ only if $s'$ is equal to $s-1, s$, or $s+1$. The
matrix ${\cal C}^{js\, j's'}$ is therefore block tridiagonal. 

Introducing the notation~:
$$
{\cal A}^{j j'}_{pq} \equiv \int\limits_0^1 {\rm d}r\ F_p^{nl} F_q^{n'l'}
\eqno\neqn$$
with the indices $p,q$ being either $1$ or $2$, the explicit expression
for the diagonal blocks is~:
$$
{\cal C}^{js\ j' s} = \sum\limits_{p,q} \zeta_{pq}^D {\cal A}^{j j'}_{pq}
\eqno\neqn$$
with the factors $\zeta_{pq}^D$ equal to~:
$$
\eqalign{
\zeta_{11}^D &\equiv v_{2s+1}^{(l)} v_{2s+1}^{(l')}\left[
{s(2s-1)\over(4s+1)(2s+2)(2s+1)}\right. +\cr
&\hskip 2.5cm\left.{(s+2)(2s+3)\over(4s+5)(2s+2)(2s+1)}\right] \cr
\zeta_{12}^D &\equiv v_{2s+1}^{(l)} v_{2s+1}^{(l')}\left[
-{s(2s-1)\over(8s+2)}-{(s+2)(2s+3)\over(8s+10)}\right] \cr
\zeta_{21}^D &\equiv \zeta_{12}^D \cr
\zeta_{22}^D &\equiv v_{2s+1}^{(l)} v_{2s+1}^{(l')}\left[
{s(s+1)(2s+1)(2s-1)\over(8s+2)}\right.+ \cr
&\hskip 2.5cm \left.{(s+2)(s+1)(2s+3)(2s+1)\over (8s+10)} \right] \cr}
\eqno\neqn$$
The lower off-diagonal blocks are the transpose of the upper off-diagonal
blocks~: ${\cal C}^{js\ j' s-1} = {\cal C}^{j' s-1\ js}$. Therefore
only expressions for the upper off-diagonal blocks are given here~:
$$
{\cal C}^{js\ j' s-1} = \sum\limits_{p,q} \zeta_{pq}^O {\cal A}^{j j'}_{pq}
\eqno\neqn$$
with the factors $\zeta_{pq}^O$ equal to~:
$$
\eqalign{
\zeta_{11}^O &\equiv v_{2s+1}^{(l)} v_{2s-1}^{(l')}{-1\over 8s+2} \cr
\zeta_{12}^O &\equiv v_{2s+1}^{(l)} v_{2s-1}^{(l')}{s(2s-1)\over 8s+2} \cr
\zeta_{21}^O &\equiv v_{2s+1}^{(l)} v_{2s-1}^{(l')}{(s+1)(2s+1)\over 8s+2} \cr
\zeta_{22}^O &\equiv v_{2s+1}^{(l)} v_{2s-1}^{(l')}{-s(s+1)(2s+1)(2s-1)
\over 8s+2} \cr}
\eqno\neqn$$
Each block has a number of elements equal to $N_{mul}\times N_{mul}$ with
$N_{mul}$ the number of multiplets in the mode-set. The number of such
blocks is equal to the number of a-coefficients available, which for
currently available mode-sets is around 20. When compared with a direct 
2-D SOLA inversion this procedure reduces the computer storage requirements 
by a factor which is roughly equal to the number of available a-coefficients 
($\sim 20$), and the CPU-time required for inversion of the matrix is 
reduced by a factor which is roughly equal to the square of that ($\sim 400$).

\titlea{Conclusions}

In this paper it is shown that the process of fitting Ritzwoller-Lavely 
type a-coefficients to ridges in the $(l,\nu )$-diagram leads to 
rotational mode kernels for the a-coefficients which can be expressed in 
closed form, as functions of associated Legendre polynomials. These 
explicit expressions for the kernels facilitate inversions for solar 
rotation as a function of radius and latitude, since it clearly is not 
necessary to apply the iterative procedure outlined in Appendix A 
of SCDT to the splittings mode kernels in order to obtain the a-coefficient 
kernels.
For convenience the results that are relevant for 2-dimensional inversions 
for solar rotation from a-coefficients are collected below. The 
equivalent expression of Eq. \Drotrel) is~:
$$
2\pi a_{nl\, 2s+1}\ =\ \int\limits_{0}^{1} {\rm d} r \int\limits_{-1}^1 
{\rm d}u\, K_{nl\, s}(r, u) \Omega (r, u)
\eqno\neqn
$$
where the `mode kernels' are now given by~:
$$
K_{nl\, s} (r, u)\ \equiv\ F_1^{nl}(r) G_1^{(l)\, 2s+1}(u)\;+\;
F_2^{nl}(r) G_2^{(l)\, 2s+1}(u)
\eqno\neqn
$$
with~:
$$
\eqalign{
F^{nl}_1\ &\equiv\ \rho(r) r^2 \left[ \xi_{nl}^2(r) - {2\over L} \xi_{nl}(r)
\eta_{nl} (r) + \eta_{nl}^2(r) \right] / I_{nl} \cr
F^{nl}_2\ &\equiv\ \rho(r) r^2 \left[ L^{-2} \eta_{nl}^2(r) \right]
/ I_{nl}
\cr}
\eqno\neqn
$$
and~:
$$
\eqalign{
G_1^{(l)\, 2s+1} &\equiv -{1\over 2}{(4s+3) v^{(l)}_{2s+1}\over 
(2s+2)(2s+1)} (1-u^2)^{1\over 2} P_{2s+1}^1 (u) \cr
G_2^{(l)\, 2s+1} &\equiv {1\over 4}v^{(l)}_{2s+1} (4s+3) (1-u^2)^{1\over 2}
P_{2s+1}^1 (u) \cr}
\eqno\neqn$$
The factors $v^{(l)}_{2s+1}$ are given by Eq. \vnorfac). The latitudinal
functions satisfy a normalization slightly different from that expressed
in condition \GLMnor)~:
\eqnam\GLSnor{GLSnor}
$$
\int\limits_{-1}^{1} {\rm d} u\, G_1^{(l)\, 2s+1}(u) = \delta_{s0} = - 
\int\limits_{-1}^{1} {\rm d} u\, G_2^{(l)\, 2s+1}(u)
\eqno\neqn
$$
With these equations a 2-D inverse problem is defined which can be solved 
using methods identical to those used to invert for the solar rotation 
rate from individual frequency splittings. 

Furthermore it is shown that SOLA inversions for 2-D rotation
can be speeded up considerably because the matrix to be inverted in the
SOLA method is shown to be symmetric block tridiagonal.

\acknow{The Theoretical Astrophysics Center is a collaboration 
between Copenhagen University and Aarhus University and is funded by 
Danmarks Grundforskningsfonden. }

\medskip
\begref{References}

\ref
Abramowitz M., Stegun I.A., 1972, 
Handbook of Mathematical Functions. 
Dover, New York, 
p. 1006
\ref
Gradshteyn I.S., Ryzhik I.M., 1994,
Table of Integrals, Series, and Products. $5^{th}$ Ed.
Acad. Press, San Diego, 
pp. 804, 808, 947, 1022, 1033
\ref
Merzbacher E., 1970,
Quantum Mechanics. $2^{nd}$ Ed.
Wiley, New York,
pp. 389-396
\ref
Pijpers F.P., Thompson M.J., 1994,
{A\&A}
281, 231
\ref
Pijpers F.P., Thompson M.J., 1996,
{MNRAS}
279, 498
\ref
Ritzwoller M.H., Lavely E.M., 1991,
{ApJ}
369, 557
\ref
Schou J., Christensen-Dalsgaard J., Thompson M.J., 1994,
{ApJ}
433, 389 (SCDT)

\endref

\bye